\documentclass{ws-procs9x6}

\def\citebk#1{[\hspace{0.9mm}\raisebox{-1.85mm}[0mm][0mm]
  {\Large\cite{#1}}\hspace{-0.1mm}]}

\newcommand{\be}{\begin{equation}}
\newcommand{\ee}{\end{equation}}
\newcommand{\bea}{\begin{eqnarray}}
\newcommand{\eea}{\end{eqnarray}}
\begin{document}

\title{QCD Sum Rules - a Working Tool for Hadronic Physics}

\author{Alexander Khodjamirian $^{*)}$  }

\address{ Institut f\"ur Theoretische Teilchenphysik, 
Universit\"at Karlsruhe,\\  
D-76128 Karlsruhe, Germany}


\maketitle

\abstracts{QCD sum rules are overviewed 
with an emphasize on the practical applications 
of this method to the physics of light and heavy hadrons.}

\marginpar{
\vspace{-8cm}
\hspace{6cm}
%\begin{flushright}
TTP02-23\\
%\end{flushright}
}

\begin{center}
{\it talk given at Continuous Advances in QCD 2002/ ARKADYFEST 
(honoring the 60th birthday of Arkady Vainshtein),\\ Minneapolis, 
Minnesota, 17-23 May 2002.
}
\end{center}

\footnotetext{ $^{*)}$ on leave from
Yerevan Physics Institute, 375036 Yerevan, Armenia}

\section{Introduction}
Imagine a big birthday cake for Arkady Vainshtein, each candle 
on that cake corresponding to one of his outstanding contributions 
to the modern particle theory. I think, a very bright and
illuminating candle should then mark QCD sum rules.

The renown papers introducing QCD sum rules \citebk{SVZ} have been 
published by Shifman, Vainshtein and Zakharov in 1979. 
The method, known also under the nickname of 
SVZ or ITEP sum rules, very soon became quite popular in the particle 
theory community, especially  in Russia. 
Not only experienced theorists, but also many students of that time
contributed to the development of this field with important results. 
It was indeed a lot of fun 
to start with an explicit QCD calculation in terms of quark-gluon 
Feynman diagrams and end up estimating dynamical characteristics 
of real hadrons. The flexibility and universality of the sum rule
method allowed one to go from one interesting problem to another,
describing, in the same framework, very different hadronic
objects, from pions and nucleons to charmonium  and $B$ mesons.  
Nowadays, QCD sum rules are still being actively used, providing many important 
applications and representing an important branch 
on the evolution tree of approximate QCD methods.  

In  this short overview, I start, in Sect.~2, from
explaining the basic idea of sum rules which is rooted in quantum mechanics.
After that, in Sect.~3, I outline the SVZ sum rule 
derivation in QCD. Some important applications and extensions
of the method are listed in Sect.~4.
Furthermore, in Sect.~5 I demonstrate how  QCD sum rules 
are used to calculate the soft contributions to the pion 
form factor. The light-cone version of sum rules is introduced.
Many interesting applications of QCD sum rules 
remain outside this survey, some of them can be found 
in recent reviews \citebk{SRrevs,CK}.

\section{SVZ sum rules in quantum mechanics}

To grasp the basic idea of the QCD sum rule method 
it is sufficient to consider a dynamical system much simpler than QCD,
that is quantum mechanics of a nonrelativistic particle in the potential
$V(r)$. The latter has to be smooth enough at small distances and confining 
at large distances. The spherically-symmetrical harmonic oscillator 
$V(r)=m\omega^2 r^2/2$ is a good example. Evidently, having defined the potential, 
one is able to solve the problem {\em exactly} e.g., by means of the 
Schr\"odinger equation, $H\psi_n(\vec{x})=E_n\psi_n(\vec{r})$, 
with the Hamiltonian $H=\vec{p}~^2/2m+V(r)$, obtaining the wave functions 
$\psi_n(\vec{r})$ and energies $E_n$ of all eigenstates, $n=0,1,...$ .

As demonstrated in \citebk{NSVZ}, it is possible to use an alternative procedure 
allowing one to calculate {\em approximately} the energy $E_0$
and  the wave function at  zero,  
$\psi_0(0)$, of the lowest level.  
The starting object is the time-evolution operator, or 
the Green's function of the particle $G(\vec{x}_2,\vec{x}_1;t_2-t_1)$,
taken at $\vec{x}_1=\vec{x}_2=0$ and written in terms of the standard 
spectral representation:
\begin{equation}
G(\vec{x}_2=0,\vec{x}_1=0; t_2-t_1)=\sum\limits_{n=0}^{\infty}|\psi_n(0)|^2 e^{-iE_n(t_2-t_1)}\,.
\label{eq:Green}
\end{equation}
Performing an analytical continuation of the time variable
to imaginary values: $t_2-t_1\to -i\tau$, one transforms
Eq.~(\ref{eq:Green})  into a sum over decreasing exponents:
\begin{equation}
G(0,0; -i\tau)\equiv M(\tau)=
\sum\limits_{n=0}^{\infty}
|\psi_n(0)|^2 
e^{-E_n \tau}\,.
\label{eq:Mtau}
\end{equation}

The function $M(\tau)$  has a {\em dual} nature
depending on the region of the variable $\tau$. 
At small $\tau$, the perturbative expansion
for $M(\tau)$ is valid, and it is sufficient to retain a few first terms: 
\begin{equation}
M^{pert}(\tau)= M^{free}(\tau)\left(1-4m\int\limits_0^{\infty}r dr  V(r)
e^{-2mr^2/\tau}+O(V^2)+...\right)\,,
\label{eq:Mpert}
\end{equation}
where 
$
M^{free}(\tau)=(\frac{m}{2\pi\tau})^{3/2}
$
is the Green's function of the free particle motion.
Using QCD terminology, we may call the behavior of $M(\tau)$ at
small $\tau$ ``asymptotically free''
having in  mind that it is approximated 
by a universal, interaction-free particle motion. 
Equating (\ref{eq:Mtau}) and (\ref{eq:Mpert}) we obtain
\be
\sum\limits_{n=0}^{\infty}
|\psi_n(0)|^2 
e^{-E_n \tau}\simeq M^{pert}(\tau)\,,
\label{eq:sr}
\ee
a typical {\em sum rule} which is valid at small $\tau$,
relating the sum over the bound-state contributions
to the result of the perturbative expansion. 
Note that the latter includes certain ``nonperturbative''
or ``long-distance'' effects too, namely the subleading terms containing the interaction
potential $V$.

At large $\tau$ one has a completely different picture. In the 
spectral representation (\ref{eq:Mtau}) the entire sum over excited
levels dies away exponentially
with respect to the lowest level contribution:
\begin{equation}
\lim\limits_{\tau \to \infty} M(\tau)=|\psi_0(0)|^2e^{-E_0\tau}\,.
\label{eq:largetau}
\end{equation}
Thus, at large $\tau$  one encounters a typical ``confinement'' regime, because 
the lowest level parameters determining $M(\tau)$ essentially 
depend on the long-distance dynamics (in this case determined by  $V(r)$ at large $r$).

%%%%%%%%%%%%%%%%%%%%%%%%%%%%%%%%%%%%%%%%%%%%%%%FIG.1%%%
\begin{figure}[hb]
%\epsfxsize=12cm   %width of figure - will enlarge/reduce the figures
%\epsfbox{figosc.ps}
%\figurebox{2cm}{3cm}{} %to have a box alone
\centerline{\epsfxsize=4.0in\epsfbox{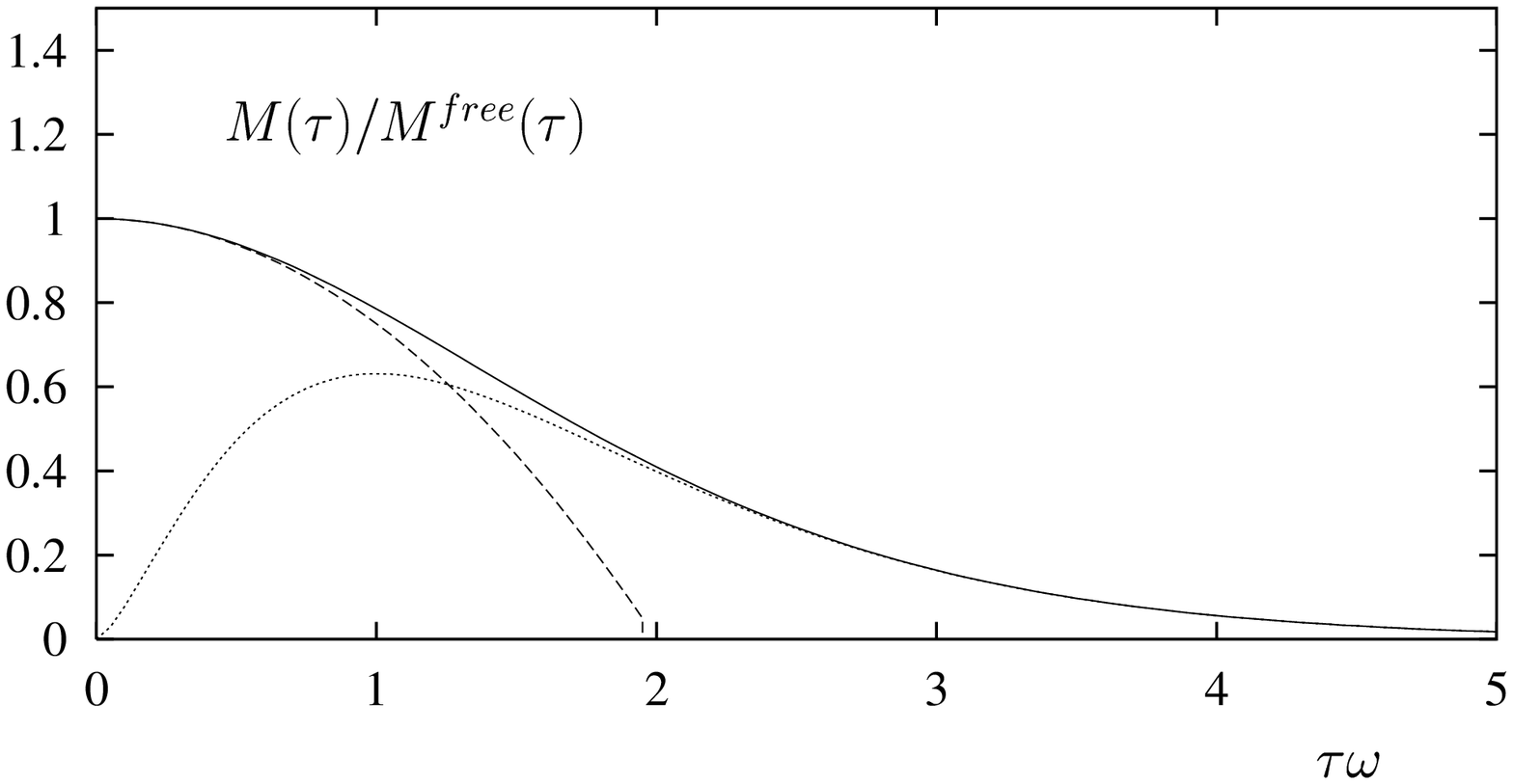}}  
\caption{The analytically continued Green's function $M(\tau)$ for a
particle in the oscillator $V(r)=m\omega^2 r^2/2$, normalized to
$M^{free}(\tau)$ and plotted as a function of $\tau \omega$. The exact
solution (solid) is compared with the perturbative calculation including
the first order in $V$ correction (dashed) and with the 
contribution of the lowest bound state (dotted).} 
\end{figure}
%%%%%%%%%%%%%%%%%%%%%%%%%%%%%%%%%%%%%%%%%%%%%%%%%%
An important observation made in \citebk{NSVZ}  is that  
at intermediate values of $\tau$ both descriptions 
(\ref{eq:Mpert}) and  (\ref{eq:largetau})
are approximately valid (see Fig.~1). It is therefore possible  
to retain only the lowest-level contribution in the sum rule (\ref{eq:sr})
allowing one  to estimate both $E_0$ and $|\psi_0(0)|$, without actually 
solving the Schr\"odinger equation. 

To further improve the quality of this determination, 
$M^{pert}(\tau)$  can be rewritten in a form of the integral 
\begin{equation}
M^{pert}(\tau)= \int\limits_0^\infty \rho^{pert}(E)e^{-E\tau}dE\,,
\label{dual}
\end{equation}
resembling the spectral representation, so that the positive
definite function $ \rho^{pert}(E)$ can be called perturbative spectral density.
The integral in Eq.~(\ref{dual}) is then splitted into two parts, introducing some
threshold energy $E_{th}>E_0$ and the 
sum over all excited states $n\geq 1$ in Eq.~(\ref{eq:sr})
is approximated by an integral over  $ \rho^{pert}(E)$
starting from this threshold:
\begin{equation}
\sum\limits_{n=1}^{\infty}|\psi_n(0)|^2 
e^{-E_n \tau} \simeq
\int\limits_{E_{th}}^\infty \rho^{pert}(E)e^{-E\tau}dE\,.
\label{eq:dual1}
\end{equation}
The latter equation can be called a ``duality'' relation
having in mind duality between the asymptotic-freedom
regime and the spectral sum.
The integral (\ref{eq:dual1}) is then subtracted from both sides of 
Eq.~(\ref{eq:sr}) leading to the sum rule for the lowest level:
\begin{equation}
|\psi_0(0)|^2e^{-E_0\tau}= \int\limits_0^{E_{th}}\rho^{pert}(E)e^{-E\tau}dE\,.
\end{equation}

Note that one could make use of the sum rule relations 
similar to Eq.~(\ref{eq:sr}) in an opposite way.
Imagine that the interaction potential is unknown but 
we have a possibility to measure, for a set of  low levels, their   
wave functions at zero separations and  energies experimentally. 
The sum rule (\ref{eq:sr}) could then be used to extract or at least
constrain the potential $V(r)$.

Interestingly,  quantum mechanics may also serve as a model for 
more complicated patterns of nonperturbative interactions, in which case
sum rules have to be treated with care. An example presented 
in \citebk{NSVZalike} is a potential containing two terms:
$V(r)=V_0[(r/r_4)^4+(r/r_{11})^{11}]$ with $r_{11}\ll r_4$, that is, 
a sharp short-range confining potential combined with a broader one. 
At $\tau\to 0$, in the perturbative
expansion of $M(\tau)$ the correction due to the second term in the potential is much
smaller than the correction associated with the first term.  
However, if one ignores the ``nonperturbative effect''
related to the short-distance scale $r_{11}$, 
the resulting sum rule simply reproduces the lowest level in the
potential $V(r)=V_0(r/r_4)^4$. In reality, the physical picture 
is quite different because it is the $\sim r^{11}$  part of the potential
which mainly determines the formation of bound states. 
The important lesson drawn from this example
is: if for some reason a short-distance nonperturbative effect 
is missing and/or ignored, the sum rule does not work (or, in other words,
duality is violated).

Interestingly, the sum rule approach in quantum mechanics 
can be generalized to  calculate more complicated characteristics
such as  the amplitudes of electric-dipole  transitions between the lowest $S$ and $P$
levels in a given nonrelativistic potential \citebk{KM}.  
One has to construct a three-point correlation function: 
\bea
\widetilde{M}(\tau_1,\tau_2)= \Big\{\!\!\int dt_3 d\vec{x}_3
\frac{\partial}{\partial|\vec{x}_2|}G(\vec{x}_2,\vec{x}_3;-i\tau_2-t_3)
\nonumber
\\
\times (\vec{e}\cdot\vec{v}_3)
G(\vec{x}_3,\vec{x}_1;t_3-(-i\tau_1))\Big \}_{\vec{x}_{1,2}=0}\,,
\label{corr3}
\eea
where $\vec{v}=i(H\vec{x}-\vec{x}H)$ is the
quantum-mechanical velocity operator, and $\vec{e}\cdot\vec{v}$ 
is the operator corresponding to the dipole radiation of a photon 
with polarization $\vec{e}$. The correlator (\ref{corr3}) corresponds 
to the propagation of a particle in $P$ wave (below threshold or in
imaginary time) from point 2 to point 3 where a dipole photon is
radiated and then further propagation to point 1 in $S$ wave. 
Calculating Eq.~(\ref{corr3}) perturbatively and matching it to 
the double spectral sum over $P$ and $S$ levels, one gets a sum rule 
which, at intermediate values of  the two variables $\tau_{1,2}$ 
is well approximated by the contributions of 
the three lowest E1 transition amplitudes ($1P\to 1S,2S\to 1P,2P\to 2S$).

The sum rule approach considered here is, of course not very important 
for quantum mechanics itself, but as we shall see in a moment, serves 
as a very convenient prototype for an analogous method 
in QCD, in the theory where no exact solution is so far available.

\section{SVZ Sum rules in QCD}

We now move from the safe haven of nonrelativistic quantum mechanics to 
QCD, a complicated theory with a rich pattern of
quark-gluon and gluon-gluon interactions. At short distances, due to  
asymptotic freedom the theory can still be resolved. One considers a quasi-free quark propagation
with calculable perturbative corrections. However, at large distances, 
$r\sim 1/\Lambda_{QCD}$, the QCD perturbation theory becomes
inapplicable and the confinement  phenomenon takes over, driven by  
the quark-gluon fluctuations in the QCD vacuum. As a result, 
quarks build coherent bound states, hadrons.
In general, it is not possible to describe QCD interactions with a potential.
Nevertheless, qualitatively, the pattern of quark-antiquark forces
in QCD,  with asymptotic freedom at small distances
and formation of bound states at large distances,  is very similar to 
the quantum-mechanical motion in the confining, oscillator-type
potential considered in the previous section.
It is therefore not surprising that sum rules \citebk{SVZ} analogous 
to the quantum-mechanical ones exist also
in QCD. 

The starting object in QCD analogous to the Green's function $G(0,0,t)$ is the
correlation function describing an evolution of a colorless
quark-antiquark pair emitted and absorbed by external 
currents. A ``classical example'' considered in \citebk{SVZ} is  
the  correlation of two $j^{\rho}_\mu=(\bar{u}\gamma_\mu u -\bar{d}\gamma_\mu d)/2$
quark currents with the $\rho^0$ meson quantum numbers (isospin 1, $J^P=1^{-}$):
\be
\Pi_{\mu\nu}(q)=i\int d^4x e^{iqx}
\langle 0 | T\{j^{\rho}_\mu(x),j^{\rho}_\nu(0)\}|0\rangle\,.
\label{eq:pimunu}
\ee
The  dispersion relation (K\"allen-Lehmann representation) 
for this correlation function contains a sum over all intermediate
hadronic states, a direct analog of the spectral representation (\ref{eq:Green})
:
\be
\Pi_{\mu\nu}(q)=\sum\limits_h\frac{\langle 0 |j^{\rho}_\mu|h \rangle
\langle h |  j^{\rho}_\nu |0 \rangle}{m_h^2-q^2} +\mbox{subtractions}\,.
\label{eq:disp}
\ee
Note that, for brevity, I wrote the above relation in a very schematic
way, including the excited $\rho$ resonances and the continuum states with $\rho$ 
quantum numbers in one discrete sum. 

The Borel transformation, $\hat{B}\{1/(m_h^2\!-\!q^2)\}\!\!\to\!\! \exp(-m_h^2/M^2)$,
converts the hadronic representation (\ref{eq:disp}) into a sum over decreasing exponents, 
\be 
\hat{B}\Pi_{\mu\nu}=
\sum\limits_{h}\langle 0 |j^{\rho}_\mu|h \rangle
\langle h |j^{\rho}_\nu |0 \rangle e^{-m_h^2/M^2}\,,
\label{eq:borel}
\ee
i.e., the inverse Borel variable $1/M^2$ plays essentially
the same role as the auxiliary variable $\tau$ in the quantum-mechanical
case. Another very important virtue of the Borel transformation is that 
it kills subtraction terms in the dispersion relation.

At large spacelike momentum transfers $q^2<0$, $Q^2\equiv
-q^2 \gg \Lambda_{QCD}^2$ (corresponding to large $M\gg \Lambda_{QCD}$ after Borel
transformation) the quark-antiquark propagation
described by the correlation function (\ref{eq:pimunu})
is highly virtual, the characteristic times/distances being 
$x_0 \sim |\vec{x}|\sim 1/\sqrt{Q^2}$. One can then benefit from asymptotic
freedom and calculate the correlation function in this region
perturbatively. The corresponding diagrams up to $O(\alpha_s)$ are depicted in Fig.~2.

As first realized in \citebk{SVZ}, there are additional
important effects due to the interactions with the vacuum
quark and gluon fields. The latter  have 
typically long-distance ($\sim \Lambda_{QCD}$) scales and, in first approximation, 
can be replaced by static fields, the {\em vacuum condensates}. 
An adequate framework to include these effects in the correlation
function was developed  in a form of the
Wilson operator product expansion (OPE). 
The Borel transformed answer 
for the correlation function (\ref{eq:pimunu}) reads:
\be
\hat{B}\Pi_{\mu\nu}^{OPE}= \hat{B}\Pi^{pert}_{\mu\nu}+
\sum\limits_{d=3,4,...} \hat{B}C_{\mu\nu}^d\langle 0|O_d |0\rangle\,,
\label{OPE}
\ee
where the first term on the r.h.s. corresponds to the perturbative diagrams
in Fig.~2, whereas the sum contains 
the contributions of vacuum condensates, ordered by their dimension
$d$. Diagrammatically, these contributions are  depicted in Fig.~3.
%%%%%%%%%%%%%%%%%%%%%%%%%%%%FIGs.23%%%%%%%%%%%%%%%%%%
\begin{figure}[h]
%\epsfxsize=12cm   %width of figure - will enlarge/reduce the figures
%\epsfbox{figAF1.ps}
%\figurebox{2cm}{3cm}{} %to have a box alone
\centerline{\epsfxsize=4.0in\epsfbox{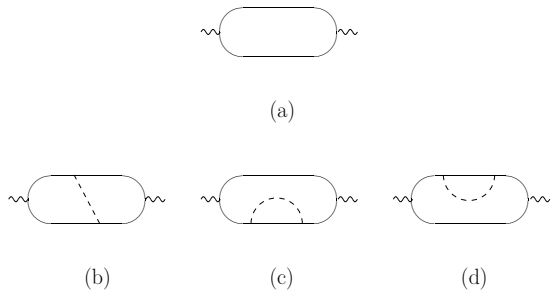}}  
\caption{Diagrams determining the perturbative part of the correlation
function (\ref{eq:pimunu}): the free-quark loop (a) and the $O(\alpha_s)$ 
corrections (b,c,d). Solid lines denote quarks, dashed lines gluons,
wavy lines external currents.}
\end{figure}
\vspace{-0.5cm}
\begin{figure}[hb]
%\epsfxsize=12cm   %width of figure - will enlarge/reduce the figures
%\epsfbox{figAF1.ps}
%\figurebox{2cm}{3cm}{} %to have a box alone
\centerline{\epsfxsize=4.0in\epsfbox{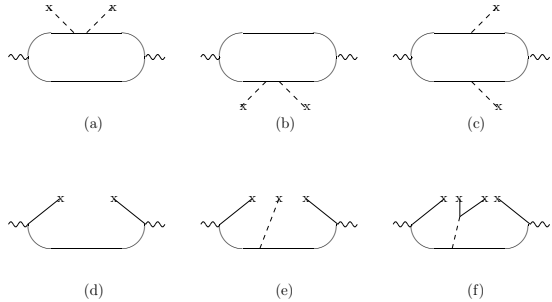}}  
\caption{Diagrams corresponding to the gluon (a,b,c,), quark (d),
quark-gluon (e) and four-quark (f) condensate contributions to the
correlation function (\ref{eq:pimunu}). The crosses denote the vacuum fields.}
\end{figure}
%%%%%%%%%%%%%%%%%%%%%%%%%%%%%%%%%%%%%%%%%%%%%%%%%%%%%%%%%%%%%%
The  terms with $d\leq 6$ contain the vacuum averages of
the operators $O_3=\bar{q}q$, 
$O_4=G^{a}_{\mu\nu}G^{a\mu\nu}$, 
$O_5=\bar{q}\sigma_{\mu\nu}(\lambda^a/2)G^{a\mu\nu}q$, 
$O_6=(\bar{q}\Gamma_rq )(\bar{q}\Gamma_s q)$, and
$O_6^{G}=f_{abc}G_{\mu\nu}^a G^{b \, \nu}_ \sigma G^{c\sigma\mu}$,
where $q=u,d,s$ are the light-quark fields, $G_{\mu\nu}^a$ is the gluon 
field strength tensor, and $\Gamma_{r,s}$ denote 
various combinations of Lorentz and color matrices.
Importantly, to compensate the growing dimension of the operator $O_d$, the
Wilson coefficients $C^d_{\mu\nu}$ contain increasing powers of $1/Q^2$.
Correspondingly, $\hat{B}C_{\mu\nu}^d$ contain  powers of $1/M^2$,
making it possible at large
$M^2$ to retain in the r.h.s. of Eq.~(\ref{OPE})
only a few first condensates. Thus, 
at $M^2\sim 1$ GeV$^2$ it is practically possible to 
neglect all operators with $d> 6$. 

Equating at large $M^2$ the hadronic representation to the 
result of the OPE calculation we obtain the desired sum rule:
\be
\sum\limits_{h}\langle 0 |j^{\rho}_\mu|h \rangle
\langle h |j^{\rho}_\nu |0 \rangle e^{-m_h^2/M^2}
=  \hat{B}\Pi^{pert}_{\mu\nu}+ \sum\limits_{d=3,4,..} \hat{B}
C_{\mu\nu}^d\langle 0|O_d |0\rangle \,.
\label{sumrule}
\ee
The explicit form of this relation is \citebk{SVZ} : 
\bea
f_\rho^2e^{-m_{\rho}^2/M^2}+ \{\mbox{excited,continuum $\rho$ states}\} 
\nonumber
\\
= M^2 \Big[\frac1{4\pi^2} \left( 1+\frac{\alpha_s(M)}{\pi}\right) 
+\frac{(m_u+m_d)\langle\bar{q}q\rangle}{M^4}  
\nonumber
\\
+\frac{1}{12}\frac{\langle \frac{\alpha_s}{\pi} G^a_{\mu\nu}G^{a\mu\nu}
\rangle}{M^4} 
-\frac{112\pi}{81} 
\frac{\alpha_s\langle \bar{q} q \rangle ^2 }{M^6}\Big]\,,
\label{SVZrho}
\eea
where the decay constant of the $\rho$ meson is defined in the standard way,
$\langle \rho^0\mid j_\nu^{\rho} \mid 0 \rangle = 
(f_\rho/\sqrt{2})m_\rho \epsilon^{(\rho)*}_\nu $.
In obtaining this relation the four-quark vacuum densities are 
factorized into a product of quark condensates. The quark-gluon
and three-gluon condensates have very small Wilson coefficients and are
neglected. The strong coupling $\alpha_s$
is taken at the scale $M$ which is the characteristic virtuality 
of the loop diagrams after the Borel transformation. A more detailed
derivation of this sum rule can be found, e.g. in the review \citebk{CK}. 
The QCD vacuum condensates were 
recently discussed in \citebk{condens}.

In full analogy with quantum mechanics, there exists a SVZ region of 
intermediate $M^2$ where the $\rho$ meson 
contribution alone saturates the l.h.s. of the sum rule (\ref{SVZrho}).
To illustrate this statement numerically, in Fig.~4 the 
experimentally measured $f_\rho$ (obtained from the $\rho^0\to e^+e^-$
width) is compared with the same hadronic parameter calculated from Eq.~(\ref{SVZrho})
where all contributions of excited and continuum states are neglected. 
One indeed observes a good agreement in the region $M^2\sim $ 1 GeV$^2$.

%%%%%%%%%%%%%%%%%%%%%%%%%%%%%%%%%%%%%%%%%%%%%%%%%%%%%
\begin{figure}[th]
%\epsfxsize=12cm   %width of figure - will enlarge/reduce the figures
%\epsfbox{figosc.ps}
%\figurebox{2cm}{3cm}{} %to have a box alone
\centerline{\epsfxsize=4.0in\epsfbox{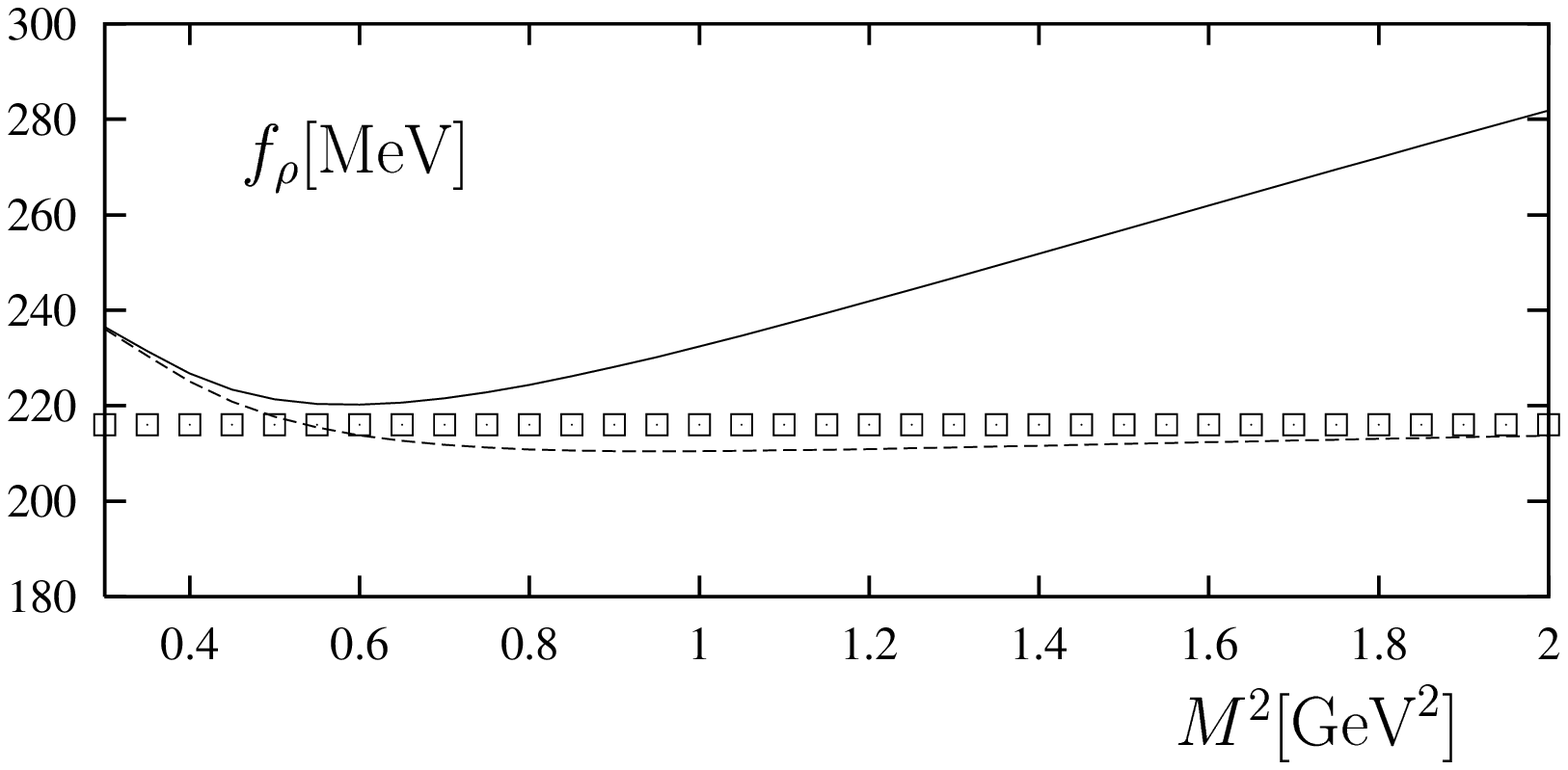}}  
\caption{The $\rho$ meson decay constant calculated from the sum rule 
(\ref{SVZrho}) neglecting all excited and continuum states (solid), 
as a function of the Borel parameter, in
comparison with the experimental value (boxes). The dashed curve 
corresponds to an improved calculation , where the sum over excited
and continuum states is estimated using quark-hadron duality
with a threshold $s_0^\rho=1.7$ GeV$^2$. }
\end{figure}

An important step to improve the sum rule (\ref{sumrule})
is to use the {\em quark-hadron duality} 
approximation. The perturbative contribution to the correlation
function (the sum of Fig.~2 diagrams) is represented
in the form of a dispersion integral and splitted into two parts:
\be
\hat{B}\Pi^{pert}_{\mu\nu}= \int\limits_0^{s_0}\rho^{pert}_{\mu\nu}e^{-s/M^2}+
\int\limits_{s_0}^\infty \rho^{pert}_{\mu\nu}e^{-s/M^2}\,.
\label{QCDdual}
\ee 
The sum over excited state and continuum contributions in Eq.~(\ref{sumrule})
is approximated
by the second integral over the perturbative spectral density
$\rho^{pert}_{\mu\nu}$. This integral is then subtracted from both parts 
of Eq.~(\ref{sumrule}).  Correspondingly Eq.~(\ref{SVZrho}) is
modified: the l.h.s. contains only the $\rho$ term ,
and, on the r.h.s., the perturbative contribution has to be 
multiplied by  a factor $(1-e^{-s_0/M^2})$. The  numerical result obtained from  
the duality improved SVZ sum rule is also shown in Fig.~4. 
Quark-hadron duality can independently be  checked for the channels with
sufficient experimental information on excited hadronic states, 
such as the $J/\psi$ and $\rho$ channels. For $\psi$ resonances one of
the first analyses of that type has been done in \citebk{GM}.

Importantly, not  all correlation functions lead to 
valid QCD sum rules. The response of the QCD vacuum to the quarks and gluons 
``injected'' by external currents crucially depends on the quantum
numbers and flavour content of the current. After all, that is the main
reason why hadrons are not alike \citebk{NSVZalike}.  In certain
channels, e.g. for the correlation functions of spin zero light-quark currents, specific  
short-distance nonperturbative effects related to instantons are present
(the so called ``direct instantons'')\citebk{NSVZalike,GI}.
These effects remain important even at comparatively large $M^2$ 
and are not accountable in a form of OPE.  The
subtle duality balance between ``quasiperturbative'' OPE  
and resonances is destroyed in such cases. Due to instantons, it is not possible, for
example, to calculate the pion parameters using correlators with 
pseudoscalar $\bar{u}\gamma_5d$ currents.
Models based on the instanton calculus have to be invoked (see ,e.g.\citebk{Shuryak}).

The QCD procedure outlined in this section has indeed many similarities 
with the sum rule derivation in quantum mechanics.
To make the analogy more transparent, in the
following table  I put together the main points
of the  two sum rule approaches: in quantum mechanics and in QCD.
%%%%%%%%%%%%%%%%%%%%%%%%%%%%%%%%%%%%%%%%%%%%%%%%%%%%%
\begin{table}[th]
{\footnotesize
\begin{tabular}{|c|c|}
\hline
{} &{}  \\[-1.5ex]
Quantum mechanics & QCD \\[1ex]
\hline
Particle in a smooth confining potential  &  quark-antiquark pair in QCD vacuum\\  
\hline
Green's function $G(0,0,t)$ & Correlation function $\Pi_{\mu\nu}(q^2)$\\[1ex]
\hline
Spectral representation & Dispersion relation in $q^2$ \\
\hline
Analytical continuation $to -i\tau$ & Borel transformation $q^2\to M^2$ \\
\hline
Perturb. expansion in powers of $V$ &  OPE (Condensate expansion) \\
\hline
matching the lowest level to $M^{pert}(\tau)$ &  matching the lowest hadron to $\hat{B}\Pi(M^2)$ \\
\hline
duality of the quasifree-motion  & quark-hadron
duality\\
and  the spectral sum &\\
\hline
extracting $V(r)$ &  extracting condensates, $m_{u,d,s,c,b}$\\
from exp. known spectral sum & from exp. known spectral density\\
\hline
sum rule does not work&  sum rules do not work \\
if the short-distance  part & in the channels
 \\
of the potential is ignored & with direct instantons \\
\hline
3-point sum rules & 3-point sum rules  \\
for E1-transition amplitudes &   for hadronic
matrix elements\\
\hline
\end{tabular}\label{tab1} }
%\end{center}
\end{table}

\section{Applying and extending the method}

\subsection{Baryons}

Following very successful applications of QCD sum rules in the mesonic
channels \citebk{SVZ}, the next essential step was to extend the
method to the baryonic sector
\citebk{barIoffe,Doschetal}. Correspondingly, 
the correlators of specially constructed quark currents with  baryon quantum numbers 
were considered. A well known example is the Ioffe current with 
the proton quantum numbers: 
\begin{equation}
J^N(x)= \epsilon_{abc} (u^{a T }(x) {\hat{C}} \gamma_\mu u^b(x) )
 \gamma_5 \gamma^\mu d^c(x)\,, 
\label{jN}
\end{equation}
where $a,b,c$ are color indices and $\hat{C}$ is the charge conjugation matrix.
From the  QCD sum rule for the correlator $\langle 0| J_N(x)J^{\dagger}_N(0)|0\rangle $ 
an approximate formula can be obtained, 
\be
m_N\simeq [-(2.0)(2\pi)^2 \langle 0| \bar{q}q |0\rangle
(\mu=1\mbox{GeV})]^{1/3}\,,
\ee 
relating the nucleon mass and the quark condensate density. Thus, QCD sum rules
unambiguously confirm the fundamental fact that $\sim 99\%$  of the baryonic mass in the Universe
is due to the vacuum condensates.

\subsection{Quark mass determination}

The sum rule relations similar to 
Eq.~(\ref{sumrule}) are widely used to  extract
the fundamental QCD parameters, not only the  condensates themselves 
but also the quark masses. One needs
sufficient experimental data on hadronic
parameters in a given channel (masses and decay constants of ground
and excited states, experimentally fitted ans\"atze for continuum
states) in order to saturate the 
hadronic part of the sum rule. 

The ratios of the light ($u,d,s$) quark masses
are predicted from the QCD chiral perturbation theory \citebk{Leutwyler}: 
\bea
\frac{m_u}{m_d}=0.553 \pm 0.043,~~~ 
%\nonumber\\
\frac{m_s}{m_d}=18.9\pm 0.8,~~~ 
\frac{2m_s}{m_u+m_d}=24.4\pm 1.5 \,,
\label{rel}
\eea
(there is also a more recent estimate $m_u/m_d= 0.46 \pm 0.09$  \citebk{Bijnens}). 
QCD sum rules offer a unique opportunity to estimate
the individual masses of $u,d,s$ quarks.
To illustrate the continuous efforts in this direction, let me 
mention one recent determination of the strange quark mass 
\citebk{JOP}, based on the correlation function 
of the derivatives of the strangeness-changing vector current
$j_\mu=\bar{s}\gamma_\mu q$, $q=u,d$:
\be
\Pi^s(q)=i\int d^4xe^{iqx}\langle 0 | T\{\partial_\mu j^\mu(x)\partial_\nu 
j^{\dagger\nu}(0)\}|0\rangle\,.
\ee
The OPE answer for $\Pi^s$ is proportional to 
$(m_s-m_q)^2\simeq m_s^2$ turning this correlator into a very convenient
object for the $m_s$ extraction. Furthermore, the recent progress in the
multiloop QCD calculations allows to reach the $O(\alpha_s^3)$ accuracy in the perturbative part
of $\Pi^s$. An updated analysis
of kaon $S$ wave scattering on $\pi,\eta,\eta'$ is used to reproduce 
the hadronic spectral density. The sum rule 
yields \citebk{JOP} for the running mass in the $\!\overline{\mbox{MS}}$
scheme: $m_s(2 \mbox{GeV})=99 \pm 16 $ MeV, in a good agreement 
with the recent lattice QCD estimates. Using the ratios (\ref{rel})
one obtains $m_{u}(2 \mbox{GeV})=2.9 \pm 0.6 $ MeV
and $m_{d}(2 \mbox{GeV})=5.2 \pm 0.9 $ MeV. 
The earlier work on predicting the light-quark masses from QCD sum rules
is summarized in \citebk{CK,Paver}.

The charmed quark mass determination was one of the first successful 
applications of the QCD sum rule approach \citebk{6auth,SVZ}. 
The correlation function of two $\bar{c}\gamma_\mu c$ currents (in other
words, the charm contribution to the photon polarization operator) was matched to its 
hadronic dispersion relation, where the imaginary  part 
is simply proportional to the $e^+e^- \to charm$ 
cross section including $\psi$  resonances
and the open charm continuum. The lowest power moments of 
this sum rule at $q^2=0$ are well suited for 
$m_c$ determination because nonpertrbative effects are extremely small. Replacing 
$c\to b$, $\psi \to \Upsilon$ and open charm by open  beauty one 
obtains analogous sum rule relations for the $b$ quark \citebk{bquark}. 
In recent years the mainstream development in the heavy quark mass determination
went in another direction, employing the higher moments which
are less sensitive to the experimental input above the
open flavour threshold. These moments, however, demand careful treatment 
of Coulomb corrections \citebk{Voloshin} which is 
only possible in the nonrelativistic QCD (the current status of this  
field is reviewed in \citebk{Hoang}). Recent precise measurements of 
the $e^+e^-\to hadrons $ cross section on one side and a substantial 
progress in  the calculation of perturbative diagrams on the other side, 
allowed to reanalyze  with a higher precision 
the low moments of the original 
SVZ sum rules for quarkonia with the following  
results \citebk{KS} for the $\overline{\mbox{MS}}$
masses: $m_c(m_c)= 1.304 \pm 0.027$ GeV, $m_b(m_b)= 4.209 \pm 0.05$ GeV.
Another subset of charmonium sum rules (higher moments at fixed large
$q^2<0$) was employed in \citebk{IoffeZ}, with a prediction for $m_c$
in  agreement with the above.

\subsection{ Calculation of the $B$ meson decay constant}

Having determined the condensates and quark masses 
from a set of experimentally proven 
QCD sum rules for light-quark and quarkonium systems
one has an exciting possibility 
to predict the unknown hadronic characteristics of  $B$ meson. 
In the amplitudes of exclusive weak $B$ decays the hadronic matrix elements 
are  multiplied by poorly known CKM parameters, such as $V_{ub}$. QCD sum rule calculations 
may therefore provide a useful hadronic input for extraction of CKM
parameters from data on exclusive $B$ decays. Importantly, the theoretical 
accuracy of the sum rule determination can be estimated by varying 
the input within allowed intervals. 

One of the most important  parameters involved in $B$ physics is the $B$ meson
decay constant $f_B$ defined via the matrix element
$\langle 0|\bar{u}i\gamma_5 b |B\rangle$.  
The calculation of $f_B$ using QCD sum rules has a long history, 
a detailed review and relevant references can be found, e,g. in
\citebk{CK,KR}, I only mention the very first papers \citebk{6auth,fB}.  
One usually employs the 
SVZ sum rule for the two-point correlator
of $\bar{b}i\gamma_5 q$ currents. I will not 
write down this sum rule explicitly. It looks very similar to the 
one for $f_\rho$ discussed in sect. 3, in a sense that the sum
rule contains a (duality subtracted) perturbative part 
and condensate terms. The expressions for the Wilson coefficients
are in this case much more complicated, especially the radiative corrections 
to the heavy-light loop diagrams. The recent essential update of the sum
rule for $f_B$ 
is worked out in \citebk{Jamin} taking into account the $O(\alpha_s^2)$ 
corrections to the heavy-light loop  
recently calculated in \citebk{ChetS}
and treating the $b$ quark mass in $\overline{\mbox{MS}}$ scheme.
The result is (for $m_b(m_b)= 4.21\pm 0.05$ GeV): 
$ f_B=210 \pm 19 $ MeV and $ f_{B_s}= 244\pm 21 $ MeV,
in a good agreement with the most recent lattice QCD determination
(including dynamical sea-quark effects).
I think, the example of $f_B$ determination demonstrates that
QCD sum rules indeed provide a reliable  analytical tool for the hadronic $B$ physics.

\subsection{Hadronic amplitudes}

To complete this short survey of 
QCD sum rule applications, it is important to
mention that this method allows to calculate 
various hadronic amplitudes involving more than one hadron.
Let me consider, as a generic example, a calculation of the hadronic matrix element 
$\langle h_f(p+q) | j | h_i(p)\rangle $ of a certain 
quark current $j$ with a momentum transfer $q$.  The convenient starting object 
in this case is the three-point correlation 
function depending on two independent 4-momenta:
\be
T_{fi} (p,q)= (i)^2 \int d^4x \; d^4y \;  
e^{-i (px + q y)} \langle 0|T\{ j_f (0) j (y)  j_i (x)\}|0 \rangle \,.
\label{eq:3point}
\ee 
As a next step, one writes down a double dispersion relation, 
in the variables $p^2$ and $(p+q)^2$ at fixed $q^2$, 
expressed in a form similar to 
Eq.~(\ref{eq:disp}):
\be
T_{fi}(p,q)=\sum\limits_{h_f}\sum\limits_{h_i}\frac{\langle 0 |j_f|h_f \rangle
\langle h_f |j|h_i\rangle 
\langle h_i |  j_i|0 \rangle}{(m_{h_f}^2-(p+q)^2)(m_{h_i}^2-p^2)} +\mbox{subtractions}\,,
\label{eq:disp3}
\ee
where the double sum includes all possible transitions between the 
states with $h_i$ and $h_f$ quantum numbers.
Two independent Borel transformations in $p^2$ and $(p+q)^{2}$  
applied to Eq.~(\ref{eq:disp3}) enhance the ground-state term containing  
the desired matrix element and allow to get rid of subtraction terms:
\be
\hat{B}_1\hat{B}_2 T_{fi}=\sum\limits_{h_f}\sum\limits_{h_i}\langle 0 |j_f|h_f \rangle
\langle h_f |j|h_i\rangle \langle h_i |  j_i|0 \rangle
e^{-m_{h_f}^2/M_2^2-m_{h_i}^2/M_1^2}\,,
\label{eq:disp3B}
\ee
where $M_1$ and $M_2$ are the Borel variables corresponding to $p^2$ and
$(p+q)^2$, respectively.
%%%%%%%%%%%%%%%%%%%%%%%%%%%%%%%%%%%%%%%%%%%%%%%%%FIG5%
\begin{figure}[tb]
%\epsfxsize=12cm   %width of figure - will enlarge/reduce the figures
%\epsfbox{figosc.ps}
%\figurebox{2cm}{3cm}{} %to have a box alone
\centerline{\epsfxsize=4.0in\epsfbox{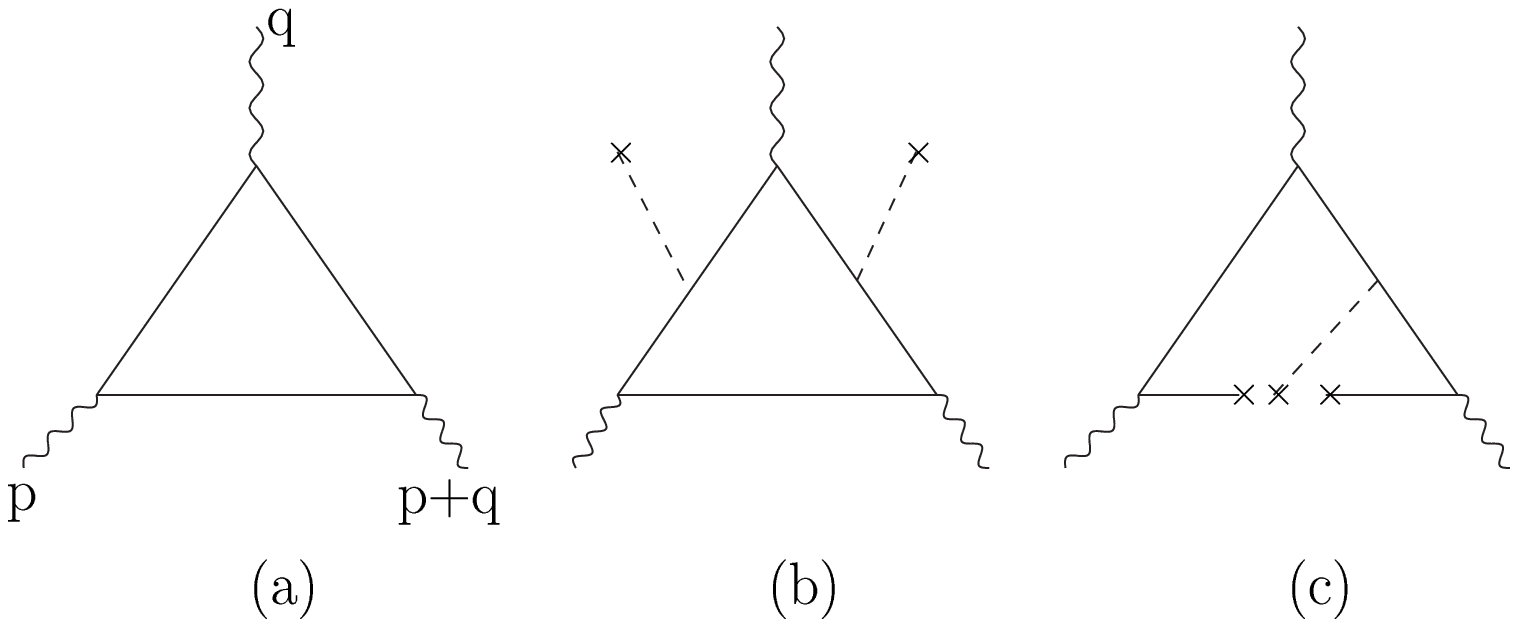}}  
\caption{ Contributions to the 3-point correlation function
  (\ref{eq:3point}): (a) perturbative, zeroth order in $\alpha_s$; 
(b)-(c) some nonperturbative corrections.}
\end{figure}
%%%%%%%%%%%%%%%%%%%%%%%%%%%%%%%%%%%%%%%%%%%%%%%%%%%
On the other hand, the correlator  (\ref{eq:3point}) can be
computed, in terms of perturbative and condensate contributions:
\be
\hat{B}_1\hat{B}_2 T^{OPE}_{fi}\!\!= 
\!\!\!\int \!\!ds\,ds'\rho^{pert}_{fi}(s,\!s^{\prime}\!,q^2)
e^{-s^{\prime}/M_2^2-s/M_1^2}
+\!\!\! \sum_{d=3,4,..} \!\!\!\hat{B}_1\hat{B}_2C^{fi}_d \langle 0|O_d|0 \rangle
\,. \label{eq:3ptOPE}
\ee 
In the above, the perturbative contribution calculated from the diagram
in Fig.~5a
is represented in a convenient form of double spectral representation
with a spectral density $\rho^{pert}_{fi}$. 
The Wilson coefficients $C^{fi}_d$ are calculated from the diagrams exemplified in
Fig.~5b,c. Importantly, one does not introduce new parameters/inputs in this
calculation, benefiting from the universality of quark/gluon
condensates. The above expansion is valid
at large spacelike external momenta: 
$|p^2|, |(p+q)^{2}| \gg \Lambda^2_{QCD}$, far from 
the hadronic thresholds in the corresponding channels. 
Accordingly, the squared momentum transfer  is also kept
large, $Q^2=-q^2\gg \Lambda^2_{QCD}$, 
at least for the currents containing light quarks \footnote{
Hadronic matrix elements at $q^2=0$, e.g., the nucleon magnetic moment 
can also be calculated within QCD sum rule approach, 
using the external (background) field technique 
\citebk{external}, with additional vacuum condensates induced
by external, non-QCD fields.}.
Equating the hadronic dispersion relation to the OPE result at large
$M_1^2,M_2^2$ and 
invoking quark-hadron duality one obtains the sum rule 
for the matrix element:
\bea
f_i f_f \langle h_f |j|h_i\rangle e^{-m_{h_f}^2/M_2^2-m_{h_i}^2/M_1^2}
&=& \int\limits_{R(s_0,s^{\prime}_0)}ds~ds'\rho^{pert}_{fi}(s,s^{\prime},q^2)
e^{-s^{\prime}/M_2^2-s/M_1^2}
\nonumber\\
&+&\sum\limits_d \hat{B_1}\hat{B_2}C^d_{fi}\langle
0|O_d|0 \rangle\,,
\label{3pointsr}
\eea
where 
$f_i=\langle h_i |j_i|0\rangle$, $f_f=\langle 0 |j_f|h_f\rangle$
are the decay constants of the initial and final hadrons.
The latter are calculable from two-point sum rules, or simply known
from experiment. In the above, 
$R(s_0,s^{\prime}_0)$ is the quark-hadron duality domain in
the $s,s^{\prime}$  plane, $s_0,s^{\prime}_0$ are 
the corresponding thresholds.
Using 3-point correlators, the sum rules for charmonium radiative 
transitions have been derived in \citebk{charmonium}.  
Another important application  \citebk{pion3p} is the pion 
e.m. form factor discussed in more detail in the next section.

\section { QCD sum rules and the pion form factor}

One of the celebrated study objects in hadronic physics is 
the pion electromagnetic form factor $F_\pi(q^2)$ 
determining the pion  matrix element 
$
\langle \pi(p+q)| j^{em}_\mu | \pi(p)\rangle =
F_\pi(q^2) (2p+q)_\mu   
$
of the quark e.m. current
$
j_\mu^{em}=e_u \bar u \gamma_\mu u + e_d \bar d \gamma_\mu d $.

At very large values of the spacelike momentum transfer $Q^2\equiv -q^2
\to \infty$ the form factor is
determined
by the perturbative QCD factorization \citebk{asympt}: 
\be
F_\pi(Q^2)= 
\frac{8\pi\alpha_sf_\pi^2}{9Q^2}\left |\int\limits_0^1 du\frac{\varphi_\pi(u)}{1-u}\right |^2\,,
\label{asympt1}
\ee
obtained by the convolution of distribution amplitudes
(DA) $\varphi_\pi(u)$ 
of the initial and final pions (see the definition below) 
with the $O(\alpha_s)$ quark-gluon hard-scattering amplitude.
At finite $Q^2$, the major problem is to estimate the ``soft'',
$O(\alpha_s^0/Q^4)$ part of this form factor. It corresponds to an overlap 
of end-point configurations of the quark-parton momenta in the initial and
final pions, so that the large momentum is transferred 
without a hard gluon exchange (the so called Feynman mechanism).

The first model-independent  estimate of the soft contribution  to the pion form
factor was provided by 
QCD sum rules \citebk{pion3p}. The three-point correlator 
(\ref{eq:3point}) was used, with $j$, $j_{i}$ and $j_{f}$ replaced by 
$j_\mu^{em}$, $j_{\nu5}$ and  $j_{\rho 5}^{\dagger}$, respectively, where
$j_{\nu5}=\bar{u}\gamma_\nu\gamma_5 d$ 
is the axial-vector current generating the pion state from the vacuum: 
$\langle \pi(p) |j_{\nu5}^{(\pi)}|0\rangle = -if_\pi p_\nu$.
The calculation based on OPE and condensates is valid at
sufficiently large $Q^2$, practically at $Q^2\sim$ 1 GeV$^2$. 
The resulting sum rule 
for the form factor written in the form (\ref{3pointsr}) has a rather compact
expression: 
\bea
f_\pi^2 F_\pi(Q^2)&=& \int\limits_{R(s_0^\pi)} ds \; ds^\prime 
\rho^{pert} (s, s^\prime, Q^2) e^{-{s + s^\prime \over M^{2}}} \nonumber \\
&+&{\alpha_s \over 12 \pi M^2} 
\langle G_{\mu \nu}^a G^{a \mu \nu} \rangle + 
{208 \pi\over 81 M^4} \alpha_s \langle \bar q q \rangle^2 
\left (1 + {2 Q^2 \over 13 M^2} \right ) \,,
\label{eq:3ptFpi}
\eea
where the perturbative spectral density is  
\be
\rho^{pert} (s, s^\prime, Q^2)={3 Q^4 \over 4 \pi^2} {1 \over \lambda^{7/2}}
\big[3 \lambda(\sigma+Q^2)(\sigma+2 Q^2) - \lambda^2 - 5 Q^2 (\sigma + Q^2)^3 
\big] \,,
\ee
with $\lambda=(\sigma+Q^2)^2- 4 s s^\prime$
and $\sigma=s+s^\prime$.
In Eq.~(\ref{eq:3ptFpi}) the condensates up to $d=6$ are included,
$m_\pi=0$, $M_1=M_2=M$ and
$s_0=s_0^{\prime}=s_0^\pi\simeq 0.7$ GeV$^2$. The duality threshold is  
inferred  from the two-point sum rule for the axial-vector channel \citebk{SVZ}.
At $Q^2=1\div3$  GeV$^2$, the form factor predicted from the sum rule
 agrees with the experimental data. E.g., compare  
$F_\pi(Q^2= 1 \mbox{GeV}^2) \simeq 0.3 $ predicted from
Eq.~(\ref{eq:3ptFpi}) with the most accurate CEBAF
data \citebk{Volmer} shown in Fig.~7 below. The good agreement indicates that 
the soft mechanism is the most important one in this region and that the $O(\alpha_s)$ 
hard scattering effect which should dominate at infinitely large $Q^2$ 
is still a small correction. (The latter corresponds to 
the gluon exchanges added to  the diagram of Fig.~5a).
At large $Q^2$ the  perturbative part of the sum rule (\ref{eq:3ptFpi})
has a  $\sim 1/Q^4$ behavior, in full accordance with our expectation 
for the soft, end-point contribution to the form factor.  
However, the condensate contributions to $F_\pi(Q^2)$ are either
$Q^2$-independent or grow $\sim Q^2/M^2$. A careful look at one of the relevant diagrams 
in Fig. 5c reveals the reason of this anomalous behavior. 
Using local (static field) condensate
approximation, one implicitly neglects  the momenta of 
vacuum quark/gluon fields. The external large momentum $p$
is carried by a single quark, which, after the photon absorption,
propagates with the  momentum $p+q$, so that the contribution 
of this diagram is $q^2$ independent. Therefore, the truncated 
local condensate expansion is not an adequate approximation to reproduce
the large $Q^2$ behavior of the pion form factor.

A possibility to calculate $F_\pi(Q^2)$ including 
both soft and hard scattering effects
at large $Q^2$ \citebk{BH,BKM}, is provided by the light-cone sum rule 
(LCSR) approach
\citebk{lcsr} combining  the elements of the theory of hard exclusive processes
\citebk{asympt} with the SVZ procedure. 

One starts with introducing a vacuum-to-pion correlation function
\begin{equation}
F_{\mu\nu}(p,q) = i\int\! d^4 x \,e^{-iqx}
\langle 0| T\{j_{\mu 5}(0) j_\nu^{\rm em}(x)\} 
| \pi(p)\rangle \;,
\label{corr}
\end{equation}
where one of the pions is put on-shell, $p^2=m_\pi^2$,  
and the second one  is replaced by the generating current 
$j_{\mu5} $. 
For this correlator a dispersion relation  
is written, in full analogy with Eq.~(\ref{eq:disp}):
\be
F_{\mu\nu}(p,q)=\sum\limits_h\frac{\langle 0 |j_{\mu 5}|h \rangle
\langle h |  j_\nu^{\rm em}|\pi(p) \rangle}{m_h^2-(p+q)^2} +\mbox{subtractions}\,.
\label{eq:dispLC}
\ee
The lowest pion-state term ($h=\pi$) in the hadronic sum,
\begin{equation}
F_{\mu\nu}^{(\pi)}(p,q) =  \frac{ if_\pi F_\pi(Q^2)(p+q)_\mu (2p+q)_\nu }
{m_\pi^2 - (p+q)^2}\,, 
\label{pion}
\end{equation}
contains the desired form factor.
%%%%%%%%%%%%%%%%%%%%%%%%%%%FIG.6 %%%%%%%%%%%%%%%%%
\begin{figure}[t]
%\epsfxsize=12cm   %width of figure - will enlarge/reduce the figures
%\epsfbox{fig.ps}
%\figurebox{2cm}{3cm}{} %to have a box alone
\centerline{\epsfxsize=5.5in\epsfbox{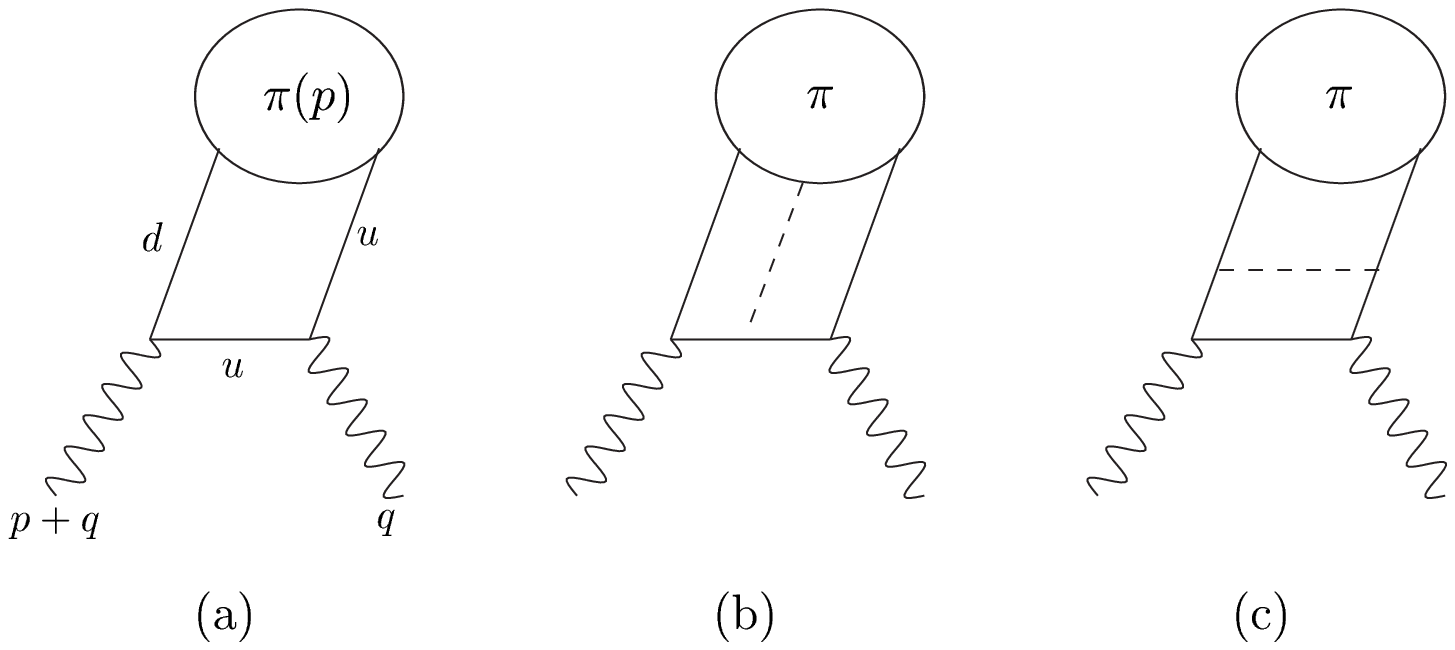}}  
\caption{ Light-cone expansion of the correlation function
(\ref{corr}).}
\end{figure}
%%%%%%%%%%%%%%%%%%%%%%%%%

At large spacelike momenta, $Q^2,|(p+q)^2| \gg \Lambda_{QCD}^2$,  
the correlation function (\ref{corr}) is dominated by small 
values of the space-time interval $x^2 $, allowing
one to expand the product of two currents 
around the light-cone $x^2=0$.
The leading-order contribution is obtained 
from the diagram in Fig.~6a and consists of two parts: (1) the
short-distance amplitude involving the virtual quark
propagating between the points $x$ and $0$,  
and (2) the vacuum-to-pion matrix element of a nonlocal quark-antiquark operator,
$\langle 0 |\bar{u}(x)\gamma_\rho\gamma_5 d(0)|\pi \rangle$.
This matrix element contains long-distance effects and is therefore
not directly calculable. On the other hand, being expanded  near $x^2=0$
it can be resolved in terms of universal distribution functions:
\bea
\langle0|\bar{u}(x)\gamma_\mu\gamma_5 d(0)|\pi(p)\rangle &=&
-ip_\mu f_\pi\int_0^1du\,e^{-iup \cdot x}
\left(\varphi_\pi (u,\mu)+x^2 g_1(u,\mu)\right)
\nonumber \\
&+&
\!f_\pi\left(\! x_\mu -\frac{x^2p_\mu}{p \cdot x}\right)\!\int_0^1
du\,e^{-iu p\cdot x}g_2(u,\mu) + ...\,,
\label{phitw4}
\eea
where the terms up to $O(x^2)$ are shown explicitly. This
expansion contains normalized light-cone distribution amplitudes (DA): 
$\varphi_\pi(u,\mu)$, $g_{1}(u,\mu)$, $g_{2}(u,\mu)$, ... , and  
the scale $\mu$ reflects the logarithmic dependence on $x^2$.
Importantly, the power moments of DA,
e.g., $M_n(\mu)=\int_0^1 du\; u^n\varphi_\pi(u,\mu)$, 
are related to the vacuum-to-pion matrix elements of local 
quark-antiquark operators with a definite {\em twist}
(dimension minus Lorentz spin). 
For that reason, $\varphi_\pi$ is called twist 2 DA,
and, correspondingly,  $g_{1}$ and $g_{2}$ are of twist 4. 
Thus, in the light-cone OPE one deals with a completely different 
pattern of  long-distance effects, 
as compared with the local OPE considered in Sect.~3.
Instead of a set of universal vacuum condensates, there is 
a set of DA for a given light meson, each of DA 
representing a series of matrix elements.

Actually, the twist 2 DA $\varphi_\pi$ was originally introduced 
in the QCD analysis of hard exclusive hadronic processes \citebk{asympt}, see 
e.g., Eq.~(\ref{asympt1}). Some of its properties
are well understood, in particular, the following expansion can be written:
\be
\varphi_\pi(u,\mu)=6u(1-u)\left(1+\sum\limits_{n=1} a_{2n}(\mu)
C_{2n}^{3/2}(2u-1)\right)\,,
\label{eq:phipi}
\ee
based on the approximate conformal symmetry of QCD with
light quarks. 
In the above, $C_{2n}$ are Gegenbauer polynomials and   
the coefficients $a_{2n}(\mu)$ 
determine the deviation of $\varphi_\pi$ from its asymptotic
form $6u(1-u)$. Due to the perturbative
evolution, $a_{2n}(\mu)$ are logarithmically 
suppressed at large $\mu$. The low-scale values of $a_{2n}$ 
(and of similar coefficients for other DA)  
have to be considered a nonperturbative input.

The  correlation function (\ref{corr}) calculated from the light-cone OPE 
represents  a convolution of the pion DA and 
short-distance (hard scattering) amplitudes:
\begin{equation}
  F_{\mu\nu}(p,q) =  2if_\pi p_\mu p_\nu \int\limits_0^1 \! du\! 
   \frac{u\varphi_\pi(u,\mu)}{(1-u)Q^2 -u(p+q)^2}+\ldots\,.
\label{ope}
\end{equation}   
For simplicity, only the leading order,  
twist 2 contribution with the relevant kinematical structure
is shown. The ellipses denote the $O(\alpha_s)$ corrections (one of the
diagrams is presented in Fig.~6c) and the higher twist contributions 
suppressed by powers of the denominator.
Physically, the higher-twist corrections take into account 
the transverse momentum of the quark-antiquark state
(e.g., the twist 4 terms in the expansion (\ref{phitw4})) and  
the contributions of higher Fock states in the pion wave function
(such as the quark-antiquark-gluon DA contributing via the diagram in
Fig.~6b.). These two effects are related via QCD equations of motion. 
More details on the pion DA of higher twists can be found in \citebk{DA}.
The factorization scale $\mu$  in Eq.~(\ref{ope})
effectively separates the large virtualities  ($>\mu^2$) 
in the hard scattering amplitude 
from the small ones $(<\mu^2)$ in the pion DA.

Equating the dispersion relation (\ref{eq:dispLC}) to the OPE result (\ref{ope})
at large spacelike $(p+q)^2$, one extracts  the form factor $F_\pi(Q^2)$ 
applying the standard elements of the QCD sum rule technique, the Borel transformation 
in the variable $(p+q)^2$ and the quark-hadron duality. The latter
reduces to a simple replacement of the lower limit in the
$u$-integration in Eq.~(\ref{ope}), $ 0 \to Q^2/(s_0^\pi+Q^2)$. 
The resulting sum rule \citebk{BH,BKM} is:
\begin{equation}
F_{\pi}(Q^2) = \!\!\!\!\!\!\int\limits_{Q^2/(s_0^\pi+Q^2)}^1 \!du\, 
\varphi_{\pi}(u,\mu) 
e^{- \frac{(1-u)Q^2}{u M^2}}
+ F_{\pi}^{(tw2,\alpha_s)}(Q^2)+F_{\pi}^{(tw4,6)}(Q^2)\,,
\label{SR1}
\end{equation} 
where the leading-order twist 2 part is shown explicitly.
The  $1/Q^4$ behavior of Eq.~(\ref{SR1}) corresponds to the 
soft end-point mechanism, provided that in the $Q^2\to \infty$ limit 
the integration region shrinks to the point $u=1$.
%%%%%%%%%%%%%%%%%%%%%%%%%%%FIG.7 %%%%%%%%%%%%%%%%%
\begin{figure}[t]
%\epsfxsize=12cm   %width of figure - will enlarge/reduce the figures
%\epsfbox{figAF1.ps}
%\figurebox{2cm}{3cm}{} %to have a box alone
\centerline{\epsfxsize=3.0in\epsfbox{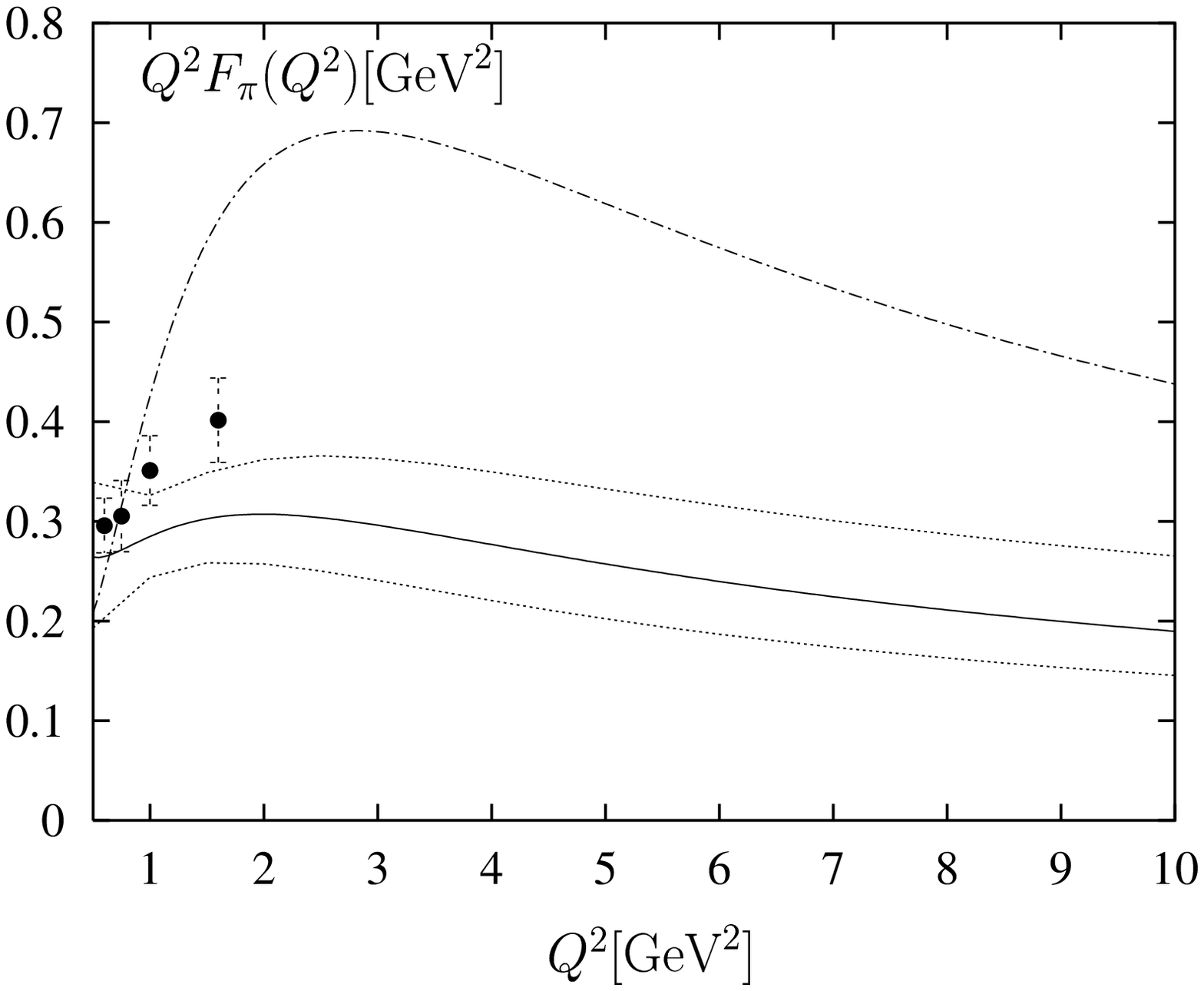}}  
\caption{The pion e.m. form factor calculated from 
LCSR in comparison with the 
CEBAF data shown with points.
The solid line corresponds to the asymptotic 
pion DA, dashed  lines indicate the estimated 
overall theoretical uncertainty ; the dash-dotted
line is calculated with the CZ model of the pion DA.}
\end{figure}
%%%%%%%%%%%%%%%%%%%%%%%%%%%%%%%%%%%%%%%%%%
The $O(\alpha_s)$ part of this sum rule 
was calculated in \citebk{BKM} and is indeed small 
at $Q^2\sim $ 1 GeV$^2$. Importantly,
in $F_\pi^{(\alpha_s)}(Q^2)$ one recovers the  $\sim 1/Q^2$ asymptotic 
term  corresponding to the hard perturbative mechanism,
with a coefficient which, in the adopted approximation, coincides
with the one in Eq.~(\ref{asympt1}). 
The higher twist contributions to the sum rule
(\ref{SR1}) manifest the same $\sim 1/Q^4$ behavior as the leading twist.
Altogether, we seem to achieve the goal. The pion form factor 
obtained from LCSR 
contains both the hard-scattering and soft (end-point) contributions,
with a proper asymptotic behavior at large $Q^2$.

The updated LCSR prediction for $F_\pi(Q^2)$  \citebk{BK} is shown in
Fig. 7. One important  practical use of this result
is to estimate/constrain the nonasymptotic coefficients $a_{2n}$
by fitting the sum rule (\ref{SR1}) to 
the experimental data on the pion form factor. However, currently
there are no sufficient data  at $Q^2>$1 GeV$^2$ to constrain 
complicated patterns of nonasymptotic coefficients. 
Considering simple ones, one finds that, e.g.  the asymptotic
DA $\varphi_\pi(u)=6u(1-u)$ is not excluded, 
whereas the CZ -model \citebk{CZ84} seems to be 
disfavored by data. Assuming that only $a_2\neq 0$ 
and neglecting all other coefficients yields \citebk{BK} the following range 
$a_2( 1\,\mbox{GeV})= 0.24 \pm 0.14\pm 0.08 $, where the first error 
reflects the estimated theoretical uncertainty and the second one 
corresponds to the experimental errors. 

Other LCSR applications to the physics of hard exclusive processes
include the $\gamma^*\gamma\pi$ form factor\citebk{AK}, 
the kaon e.m. form factor\citebk{BK}, and the first attempt
to calculate the nucleon form factors \citebk{BLMS}. 

Furthermore, an important task of LCSR is to provide $B$ physics 
with various heavy-to-light hadronic matrix elements.
In particular, the sum rule for the $B\to\pi$ form factor
can be obtained from a correlation function 
very similar to the one depicted in Fig.~6, if one 
replaces the virtual light quark  by a $b$ quark.
The calculable short-distance part will then change considerably,
but the long-distance part remains essentially the same, 
determined by the set of pion DA. 
The sum rule predictions for the $B\to\pi$ \citebk{LCSRB} 
and $B\to \rho$ \citebk{BallBraun} form factors are already used 
to extract $ |V_{ub}|$ from the widths of $B\to \pi(\rho)l \nu_l$ decays.
One can also employ LCSR to estimate  
the  hadronic amplitudes for $B\to \pi\pi$  and similar decays
beyond factorization \citebk{Bpipi}. A summary 
of the sum rule  applications to the heavy flavour physics 
can be found in \citebk{MF}.

\section{Conclusion} 
For more than twenty years, QCD(SVZ) sum rules serve as a virtual 
laboratory for studying the transition from 
short- to long-distance  QCD. The practical use of this analytical
method is twofold. 
On one hand, using sum rules for experimentally known hadronic
quantities, QCD parameters such as quark masses are extracted. On the other hand,
the sum rules are employed to predict unknown hadronic 
parameters, for example $f_B$, with a controllable accuracy.
Finally, the example of the pion form factor calculation demonstrates 
that the light-cone version of QCD sum rules has a large potential in
describing exclusive hadronic transitions.

\section*{Acknowledgments}

I am grateful to Misha Shifman, Arkady Vainshtein, and Misha Voloshin 
for hospitality and for organizing a very enjoyable and fruitful workshop. 
This work is supported by BMBF (Bundesministerium f\"ur Bildung und Forschung).

\end{document}